\documentclass{elsart}
\usepackage{amssymb}
\begin{document}

\begin{frontmatter}

\title{The universe seen at different scales}

\author[label1]{George F.R. Ellis}\ead{ellis@maths.uct.ac.za},
\author[label2]{Thomas Buchert}
\ead{buchert@theorie.physik.uni-muenchen.de}

\address[label1]{Mathematics Department, University of Cape Town,
Rondebosch 7701 \\Cape Town, South Africa}

\address[label2]{Arnold Sommerfeld Center for Theoretical Physics\\
Ludwig--Maximilians--Universit\"{a}t, Theresienstra{\ss}e 37,
80333 M\"{u}nchen, Germany}

\begin{abstract}
A large--scale smoothed--out model of
the universe ignores small--scale inhomogeneities, but the averaged effects
of those inhomogeneities may alter both observational and dynamical
relations at the larger scale. This article discusses these effects, and
comments briefly on the relation to gravitational entropy.
\end{abstract}

\end{frontmatter}

\section{Different scale descriptions: coarse--graining the gravitational field}

Any mathematical description of a physical system depends on an
\textit{ averaging scale} characterizing the nature of the
envisaged model. This averaging scale is usually hidden from view:
it is taken to be understood. Thus, when a fluid is described as a
continuum, this assumes one is using an averaging scale large
enough that the size of individual molecules is negligible. If the
averaging scale is close to molecular scale, small changes in the
position or size of the averaging volume lead to large changes in
the measured density and velocity of the matter, as individual
molecules are included or excluded from the reference volume. Then
the fluid approximation is no longer applicable; rather one is
using a detailed description of the fluid where individual
molecules are represented. Usual work referring to the fluid
density and velocity assumes a medium--size averaging scale: not
so small that molecular effects matter, but not so large that
spatial gradients in the properties of the fluid are significant
(\cite{bat67}, p.5). The actual averaging scale, or rather the
acceptable range of averaging scales, is not explicitly stated but
is in fact a key--feature underlying the description used, and
hence the effective macroscopic dynamical laws investigated.
Indeed, different types of physics (particle physics, atomic
physics, molecular physics, macroscopic physics, astrophysics)
correspond to different assumed averaging scales. Thus, instead of
referring to a density function $\varrho$, one should really refer
to a function $\varrho_{L}$: the density averaged over volumes
characterized by scale length $L$. The key--point about the fluid
approximation is that, provided this length scale is in the
appropriate domain, then its actual value does not matter; i.e.
when it is in this range, then changing $L$ by a factor of $10$,
$100$, or even much more makes no difference:\ the measured
density and average velocity will not change. But if you change
$L$ by a very large amount until outside this range, this is no
longer true. Hence, there is a range of validity
$L_{1} < L < L_{2}$ where the fluid approximation holds
\cite{bat67} and explicit mention of the associated averaging
scale may be omitted.

In electromagnetic theory, polarization effects result from a
large--scale field being applied to a medium with many microscopic
charges. The macroscopic field $E$ differs from the point--to--point
microscopic field, which acts on the individual charges because of
a fluctuating internal field $E_{i}$, the total internal field at
each point being $D=E+E_{i}$ (\cite {jac67}, p.116). Spatially
averaging, one regains the average field because the internal
field cancels out: $E=\langle D\rangle$, indeed this is how the
macroscopic field is defined (implying invariance of the
background field under averaging: $E=\langle E\rangle)$. On a microscopic
scale, however, the detailed field $D$ is the effective physical
quantity, and so is the field ``measured'' by electrons and protons
at that scale. Thus, the way different test objects respond to the
field crucially depends on their scale (a macroscopic device will
measure the averaged field).

Now, exactly the same issue arises with regard to the gravitational
field. Applications such as the solar system tests of general
relativity theory, and in particular Einstein's triumphant
prediction of light bending by the Sun, are at solar system
scales. We apply gravitational theory, however, at many other
scales: to star clusters, galaxies, clusters of galaxies, and
large--scale structures (walls and voids), as well as to black
holes (occurring at solar system and star cluster scales, and
possibly at much smaller scales).

Cosmology utilizes the largest
scale averaging envisaged in astrophysics:\ a representative scale
is assumed that is a significant fraction of the Hubble scale, and
the  cosmological velocity and density functions are defined by
averaging on such scales (\cite{ell71}, p.111). Einstein first
introduced the fluid approximation in his 1917 static universe
model, as well as a highly idealized macroscopic model of the
large--scale (smoothed) geometry of the universe. This geometrical
idealization was then canonized via Milne's \textit{cosmological
principle} (\cite{wei72}, p.408), or a somewhat more general
\textit{Copernican principle} (\cite{hawell73}, pp.134, 350);
the resulting locally isotropic, constant--curvature Robertson--Walker geometries
(\cite {hawell73}, Sect. 5.3) are nowadays taken to be a good description of the
known region of the universe. The best justification of this
assumption is the measured high degree of isotropy of the cosmic
blackbody background radiation, taken together with a Copernican
assumption (see \cite{hawell73}, pp.351--3, for the argument in
the case of exact isotropy, and \cite{stoellmaa95} or
\cite{ellvan99a}, Sect. 8.5 for the case of almost--isotropy).

However, a range of scales of description are relevant to
cosmology. There are levels of approximation in modelling the
universe, each with a hidden averaging scale. One can have a
description in which every star is represented, or every galaxy
(the stars averaged over), or only the largest scale cosmological
structures (even galaxies averaged over, as in the fluid
approximation). A typical cosmological simulation of  ``dark
matter'' gravitational clustering uses Newtonian theory and
resolves fluid elements that still contain $10^{60}$ dark matter
particles. This implicit coarse--graining can be made explicit
within a Newtonian kinetic description: introducing filtering
scales for a distribution of N self--gravitating particles in
phase space reveals that the washed out small--scale degrees of
freedom are represented by additional force terms that account for
the dynamical coupling to these degrees of freedom
\cite{buchertdominguez}; they can also be modelled by
phenomenological noise and/or stochastic forces
\cite{buchertetal99}, \cite{mabertschinger}, and can lead to
drastic qualitative changes of the system. However, this kind of
calculation would be much more difficult in a General Relativity
context.

The General Relativistic cosmological perturbation solutions used
to study structure formation embody two interacting levels: the
background (zero--order) model, almost always a Robertson--Walker
metric, and the perturbed (first--order) model representing the
growth of inhomogeneities, represented by a perturbed
Robertson--Walker metric. The question then is how do models on
two or more different scales relate to each other in Einstein's
gravitational theory \cite{ell84}. This is a difficult issue both
because of the non--linearity of Einstein's equations, and because
of the lack of a fixed background spacetime -- one of the core
features of Einstein's theory. This causes major problems in
defining suitable averaging processes as needed in studying these
processes. While there have been many analyses of this problem,
there are still issues to be resolved in relation both to
observations and dynamics, and in how this relates to
gravitational entropy and the arrow of time.

\section{Non--commutativity of averaging and observations}

The usual analysis of cosmological observations is based on the
Mattig equations relating apparent magnitude and redshift
\cite{san61}, \cite {ell71}, derived from analyzing the behaviour
of null geodesics in Robertson--Walker spacetimes. In terms of the
Sachs optical scalar equations, for hypersurface--orthogonal null
geodesics in a general spacetime, the basic equations are:
\begin{equation}
\frac{d\theta }{dv}=-R_{ab}K^{a}K^{b}-2\sigma ^{2}-\frac{1}{2}\theta ^{2}\qquad;\qquad
\frac{d\sigma _{mn}}{dv}=-E_{mn}\;\;,
\end{equation}
where $\theta $ is the rate of expansion of the null geodesics with
tangent vector $K^{a}=dx^a/dv$ and affine parameter $v$, $\sigma_{mn}$
is their shear, $R_{ab}$ the Ricci tensor, and $E_{mn}$ a matrix
of Weyl tensor components (\cite {hawell73}, p.88;
\cite{schetal92}, pp.108--9). In the idealized Robertson--Walker
case, the Weyl tensor  $C_{abcd}$ vanishes, but the Ricci tensor is non--zero,
being given via the Einstein field equations from the matter
present. Thus, $C_{abcd}\,=\,0\;\Rightarrow\; E_{mn}\,=0\,$, and the relevant
solutions are shear--free:
\begin{equation}
\sigma^{2}\,=\,0\quad\Rightarrow\quad \frac{d\theta}{dv}\,=\,
-R_{ab}K^{a}K^{b}-\frac{1}{2} \theta ^{2}\;\;.  \label{rayfrw}
\end{equation}
Integration gives the Mattig relations applicable to Friedmann
universe models (\cite{schetal92}, pp.134--7), also elegantly
obtainable from the geodesic deviation equations with vanishing
Weyl Tensor \cite{ellvan99}.

However, in the real universe, observations take place via null geodesics
lying in the empty spacetime between galaxies (you can't see the further
galaxy, if there is one in the foreground). Thus, the real situation in a
universe with no intergalactic medium (all the matter is concentrated in
galaxies) is the opposite of that above: in the region of spacetime
traversed by the geodesics, the Ricci tensor vanishes, so
\begin{equation}
\frac{d\theta }{dv}\,=\, -2\sigma ^{2}-\frac{1}{2}\theta ^{2}\;\;,
\end{equation}
but the non--zero Weyl
tensor (the tidal field caused by nearby matter) generates shear that then
causes focussing. Thus, the microscopic description of the focussing ($\sigma
\neq 0,$ $R_{ab}\,=\,0,$ $E_{mn}\neq 0$) is radically different from the
macroscopic one ($\sigma \,=\,0,$ $R_{ab}\neq 0,$ $E_{mn}\,=0$), and the area
distance--redshift relation may be expected to be different on microscopic
scales (i.e. the small solid angle bundles of null geodesics actually used
in observations of individual objects), as compared with macroscopic scales
(averaging over large solid angles).

Various proposals have been made to deal with this. The most
popular is the Dyer--Roeder distance \cite{dyeroe74,dyeroe75},
obtained by assuming only a fraction $f$ of the total mass density
is encountered by the light--rays but ignoring the shear. Thus, in
(\ref{rayfrw}) one replaces $R_{ab}K^{a}K^{b}$ by $fR_{ab}K^{a}K^{b}$
and works out the corresponding area distance
(\cite{schetal92}, pp.138--143; \cite{demetal03}). This may be a
good approximation if galaxies are embedded in a fairly uniform
intergalactic medium of dark matter, but clearly does not take
shear effects properly into account. How good it is will depend on
the nature of clustering in the universe and how the averaged
distribution impacts along the line of sight \cite{linder}.

One can approach the topic in other ways: for example by using
stochastic methods \cite{ber66}, or detailed examination of
geodesics in Swiss--Cheese universe models \cite{kan69,boukan75}. It has
been suggested that energy conservation will imply that the effect
averages out over the entire sky \cite{wei76}, but this
calculation assumed that areas of a bundle of null geodesics were
the same in the perturbed and background models, which will not be
the case when one takes the effect of caustics into account
\cite{elletal98}. Indeed, areas increase slower than in a Robertson--Walker model
in the empty spaces between matter, where the Ricci term is zero,
and faster in the high--density regions where matter is
concentrated, so one might think these effects cancel out. However,
the strongly lensed rays soon go through a caustic and emerge
highly divergent, so that areas are rapidly increasing again. It
is plausible that on average the overall effect is always an
increase in area, that is a lesser area distance than in the
smooth background model.

The potential importance of this effect is in relation to the
interpretation of the Supernova data \cite{kantho00,pynbir04,baretal05},
which is
usually taken to imply the existence of a cosmological constant or
quintessence causing acceleration of the universe at recent times \cite{linder04}.
Kantowski \cite{kan98} has obtained
analytic expressions for distance--redshift relations that have
been corrected for the effects of inhomogeneities in the density.
The values of the density parameter and cosmological constant
inferred from a given set of observations depends on the
fractional amount of matter in inhomogeneities and can
significantly differ from those obtained by using the Mattig
relations. As an example, a determination of $\Omega_0$ made by
applying the homogeneous distance--redshift relation to SN 1997ap
at z = 0.83 could be as much as 50\% lower than its true value. It
could be that the apparent acceleration term detected is at least
partly due to this optical effect:\ focussing of null geodesics is
different in a lumpy universe than in a smooth one. Clearly, this
effect needs careful investigation.

\section{Non--commutativity of averaging and dynamics}

The key--point in considering dynamical effects is that the two processes
involved in relating the field equations at different scales do not commute
\cite{ell84}. These processes are:

\begin{itemize}

\item[{\bf E}:] calculating the Einstein tensor $G_{1ab}\,:=\,R_{1ab}-\frac{1}{2}
R_{1}g_{1ab}$ from a metric tensor $g_{1ab}$, and, hence,
determining the quantity $E_{1ab}\,:=\,G_{1ab}-\kappa T_{1ab}$ for
$g_{1ab},$ where $T_{1ab}$ is the matter tensor appropriate to the
scale represented by $g_{1ab}$;

\item[{\bf A}:] averaging the metric tensor $g_{1ab}$ to produce a smoothed metric
tensor $g_{2ab}:$ $g_{2ab}=\langle g_{1ab}\rangle$ and the matter tensor $T_{1ab}$ to
produce a corresponding smoothed matter tensor $T_{2ab}:$ $T_{2ab}=\langle T_{1ab}\rangle$.

\end{itemize}

Now in general the averaging process does not commute with taking
derivatives: for a function $g$, usually $\partial_i \langle
g\rangle \; \neq\; \langle\partial_i g\rangle$ (see equations (8),
(9) below for specific examples). Furthermore the inverse metric
$g_{2}^{ab}$ (non--linearly dependent on the metric tensor
components $g_{1ab})$ is not the smoothed version of $g_{1}^{ab}.$
The resulting Christoffel terms $\Gamma_{2bc}^{a}$ are therefore
not the smoothed version of $\Gamma_{1bc}^{a},$ hence the Ricci
tensor components $R_{2ab},$ non--linearly dependent on
$\Gamma_{2bc}^{a}$, are not the smoothed versions of $R_{1ab}$.
Extra non--linearities occur in calculating the Einstein tensor
$G_{2ab}\;=\;R_{2ab}-\frac{1}{2}R_{2}g_{2ab}$ from the Ricci
tensor $R_{2ab}$. Thus, if you smooth first and then calculate the
field equations, you get a different answer than if you calculate
the field equations first and then smooth; symbolically ${\bf
A}({\bf E}(g_{1ab}))\;\neq\; {\bf E}({\bf A}(g_{1ab}))$ .

Suppose the field equations are true at the first scale: $E_{1ab}\;=\;0\;$, then
they will not be true at the second scale:\ $E_{2ab}\;:=\;G_{2ab}-\kappa
T_{2ab}\neq 0$. Thus, there will be an extra term in the equations at the
smoother scale. We can either regard it as an extra term on the left--hand--side,
\begin{equation}
G_{2ab}-E_{2ab}\;=\;\kappa\, T_{2ab}\;\;,
\end{equation}
representing a modified curvature term, or as an extra term on the right--hand--side,
\begin{equation}
G_{2ab}\;=\;\kappa \,T_{2ab}+E_{2ab}\;\;,
\end{equation}
where it is regarded as an extra contribution to the matter tensor. Which is
the more appropriate interpretation depends on the context.

Szekeres \cite{sze71} developed a polarization formulation for a
gravitational field acting in a medium, in analogy to electromagnetic
polarization. He showed that the linearized Bianchi identities for an almost
flat spacetime may be expressed in a form that is suggestive of Maxwell's
equations with magnetic monopoles. Assuming the medium to be molecular in
structure, it is shown how, on performing an averaging process on the field
quantities, the Bianchi identities must be modified by the inclusion of
polarization terms resulting from the induction of quadrupole moments on the
individual ``molecules''. A model of a medium whose molecules are harmonic
oscillators is discussed and constitutive equations are derived. This results
in the form:
\begin{equation}
E^{2ab}\;=\;Q^{abcd}_{\;\;\;\;\;\;\; ;cd}\;\;,
\end{equation}
that is $E_{2ab}$ is expressed as the double divergence of an
effective quadrupole gravitational polarization tensor $Q^{abcd}$
with suitable symmetries:
\begin{equation}
Q^{abcd}\;=\;Q^{[ab][cd]} \;=\;Q^{cdab}\;\;.
\end{equation}
Gravitational waves are demonstrated to slow down in such a medium.

The problem with such averaging procedures is that they
are not covariant. They can be defined in terms of the background
unperturbed space, usually either flat spacetime or a
Robertson--Walker geometry, and so will be adequate for linearized
calculations where the perturbed quantities can be averaged in the
background spacetime (although even here the gauge problem arises,
see below). But the procedure is inadequate for non--linear cases,
where the integral needs to be done over a generic lumpy
(non--linearly perturbed) spacetime that are not ``perturbations'' of
a high--symmetry background. However, it is precisely in these cases
that the most interesting effects will occur.

The only tensor integrals that are well--defined over a generic spacelike
surface or spacetime region (and one interesting issue is which of these one
should use) are for scalars \cite{buchert:grgdust,buchert:grgfluid},
unless one uses the bitensors associated with
Synge's world function \cite{bitensor}, based on parallel propagation along
geodesics, to compare tensors at different points in a normal neigbourhood.
The problem is that they cannot be used for averaging the metric tensor, for
it is the metric tensor itself that defines the parallel propagation used in
this process, and so is left invariant by it (since $g_{ab;c}=0$). So, one
has to devise a procedure in which either the field equations are represented
only in terms of scalars, possible for example if one takes components
relative to a covariantly uniquely defined tetrad, or else bitensors are
used to define averages of quantities other than the metric.

Zalaletdinov has taken this issue seriously, and provided the only sustained
such attempt based on bitensors \cite{zal97}. He proposes a macroscopic
description of gravitation based on a covariant spacetime averaging
procedure. The geometry of the macroscopic spacetime follows from averaging
Cartan's structure equations, leading to a definition of correlation
tensors. Macroscopic field equations (averaged Einstein equations) can be
derived in this framework. It is claimed that use of Einstein's equations
with a hydrodynamic stress--energy tensor means neglecting all gravitational
field correlations, and a system of macroscopic gravity equations is given
when the correlations are taken into consideration. This approach has
not won many adherents, but is nevertheless a systematic and coherent
attempt to set up the problem generically.

\section{Gravitational radiation}

So far, there are two main applications of dynamical averaging. The
first is to the issue of gravitational radiation. When
electromagnetic radiation is present, one can characterize it as
the high--frequency part of the electromagnetic field \cite{ell71},
and assign to it an energy density and momentum. This leads to an
energy--momentum tensor that then serves as a source of curvature
in the Einstein field equations. An obvious question is if one can
do the same for gravitational radiation:\ can one identify it
locally, and then assign to it an energy density and momentum? If so,
there should be a form of the gravitational equations where this
high--frequency part of the gravitational field acts as an effective
source of spacetime curvature. But this is a version of the
problem described above: it is just the definition of a
contribution $E_{2ab}$ to the macroscopic gravitational field due
to the fine--scale structure of the high--frequency radiation.

The problem is that gravitational radiation is only easily
determined in linearly perturbed spacetimes; in more general
spacetimes it is not easy to define the gravitational radiation
part of the curvature, except near infinity in asymptotically flat
spacetimes. Isaacson \cite{isa68a,isa68b} considered the case of
linear perturbations about flat spacetime, determining the
backreaction due to the gravitational radiation in this case. He
obtained a close analogy with the electromagnetic situation: the
`shortwave approximation' shows how the stress--energy in the
waves creates background curvature (\cite{mtw}, Sect. 35.13).
 A similar process can be applied to gravitational radiation in
 cosmological backgrounds, and
backreaction by low--frequency gravitational radiation has been
discussed by Dautcourt \cite{dautcourt}. To understand non--linear
phenomena in gravitational radiation, the possibility of solitonic
solutions and caustics should also be of concern, since these
phenomena are presumably easier to detect.

\section{Cosmology:\ Backreaction}

The second application is to understand the nature of the backreaction of
perturbations in cosmology \cite{buchert}. Unlike the gravitational
radiation case, where one averages over tensor perturbations, here one first thinks
of averaging  over scalar quantitites like the density or the rate of expansion, in order to
get control on cosmological parameters in an inhomogeneous universe model.
As long as one works with exact equations for the evolution of those fields in a given
foliation of spacetime, such an averaging procedure is covariant, e.g. for
idealized cases like dust or a perfect fluid we can work in the `covariant fluid gauge'
\cite{bruni1,bruni2}. For these cases generalized forms of Friedmann's equations
can be employed to study backreaction \cite{buchert:grgdust,buchert:grgfluid}.
As soon as we invoke explicit model assumptions, e.g. perturbation theory,
one runs directly into the gauge problem for cosmological
perturbations, as well as the covariance question mentioned above.

The gauge problem is the following: when you perturb a smooth
background cosmological metric $\bar{g}_{1ab}$  to obtain a
perturbed metric $g_{ab}\,=\, \bar{g}_{1ab}+h_{1ab}$, the inverse
relation is not unique:\ there is no agreed averaging or fitting
process that will give back a unique background metric
$\bar{g}_{ab}$ back from the ``lumpy'' metric $g_{ab}$
\cite{ellsto87}. Some other smooth metric $\bar{g}_{2ab}$ could
have been chosen as the background metric instead, leading to a
different definition of the perturbations: $h_{2ab}:=g_{ab}-\bar{g}_{2ab}$,
instead of $h_{1ab}:=g_{ab}-\bar{g}_{1ab}$. The choice of background metric
$\bar{g}_{ab}$ for a specific ``lumpy'' metric $g_{ab}$ is called a
`gauge choice'. The backreaction problem will look very different
if described in terms of different gauges.

The best way to look at this is to think of a gauge choice as a
mapping of a smooth background metric $\bar{g}_{1ab}$ into the
lumpy universe with metric $g_{ab}$ \cite{ellbru89}. At each point
in the real spacetime the density perturbation $\delta \varrho$ is
then defined by $\delta\varrho :=\varrho - \bar{\varrho}$, where
$\varrho$ is the actual density at that point, and $\bar{\varrho}$
the background density at the same point. The key--issue is the
choice of surfaces of constant time in the perturbed spacetime,
conventionally taken to represent the image of surfaces of
constant density of the background spacetime. It then becomes
clear that one can for example set the density perturbation to
zero by choosing the mapping so that the surfaces of constant
background density $\bar{\varrho}$ are the same as the surfaces of
constant real density:\ for then at each point $\varrho
=\bar{\varrho} $ $\Rightarrow \delta \varrho =0$. However, with
this choice, the fluid flow lines will not be orthogonal to the
surfaces of constant density, so there will still be a non--zero
density variation measured by comoving observers. Gauge issues
arising in treating multi--component fluids raise extra issues
because of the multiple possible choices of reference velocity
field \cite{dunsby}.

The remedy to this disconcerting behaviour is to choose gauge
invariant variables, for example a set of covariantly defined
variables that vanish in the background spacetime \cite{ellbru89}.
While many studies have been carried out for quantifying
backreaction effects in cosmology, where the smoothed--out effect
of the small--scale perturbations causes extra terms in the
Friedmann equations for the background metric, none have been done
that both fully and clearly take the gauge issue into account and
go beyond linear order. This is a key--issue waiting to be
resolved. One certainly wants to go at least to second order in
understanding the effects of non--linear perturbations, and while
linear perturbations are well--understood, there are still many
competing second order methods without a proper consensus on their
implications emerging yet.  Many of the crucial results at linear
order no longer hold, for example scalar, vector and tensor
perturbations are no longer independent
of each other at second order \cite{clarkson,lan05}, 
and then the backreaction in turn affects the perturbations
themselves \cite{marbra05}.

Isaacson's method mentioned above has been used in the
cosmological context \cite{futamase}, as has Zalaletdinov's
\cite{coletall05}. In Zalaletdinov's approach to the averaging
problem in cosmology, the Einstein field equations on cosmological
scales are modified by appropriate gravitational correlation
terms. For a spatially homogeneous and isotropic macroscopic
spacetime, the correlation tensor is of the form of a spatial
curvature term. However, it is not clear how this approach relates
to the gauge problem.

There is no doubt that interesting effects occur. How can we
design a strategy that allows both making contact with the
well--developed inventory of Friedmannian cosmology and
quantifying backreaction effects? Cosmological parameters like the
rate of expansion or the mass density are to be considered as
volume--averaged quantities, and only these can be compared with
cosmological observations. For this reason we expect that the
relevant parameters are intrinsically scale--dependent unlike the
situation in a Friedmannian cosmology. Averaging scalar
characteristics on a Riemannian spatial domain delivers the
effective dynamical sources that an observer would measure, but
although he measures within the lumpy spacetime, he -- due to a
lack of better standards -- is going to interpret his observations
within a Friedmannian fitting model. This suggests a logical
division of the averaging problem into 1) calculating averages in
the real manifold, and 2) determining the mapping between averages
in the real manifold and averages in the Friedmannian model. The
first averaging is straightforward for scalars, as we mentioned
above, and it encounters what we may call {\it non--commutativity
of averaging and time--evolution}: this is a purely kinematical
property that can be expressed, for a scalar field $\psi$, through
the rule
\begin{equation}
\label{commutation1}
\partial_t \langle \psi \rangle - \langle\partial_t \psi\rangle\;=\;
\langle\theta\psi \rangle - \langle \theta \rangle \langle\psi \rangle\;\;.
\end{equation}
The fluctuation part on the right--hand--side of this rule
produces the {\it kinematical backreaction}, which is now studied
in the context of the {\it dark energy problem} (see below). The
second ``averaging'' is more adequately thought of as a rescaling
of the tensorial geometry. A (Lagrangian) smoothing as opposed to
(Eulerian) rescaling of the metric on regional spatial domains has
been proposed by Buchert and Carfora \cite{buc02}, using a global
Ricci deformation flow for the metric. The smoothing of geometry
implies a renormalization of averaged spatial variables,
determining the effective cosmological parameters as they appear
in the Friedmannian fitting model. Two effects that quantify the
difference between background and real parameters were identified:
{\it curvature backreaction} and {\it volume effect}
\cite{buccar03}. Both are the result of an inherent {\it
non--commutativity of averaging and spatial rescaling}. In this
way we look at the averaging problem in two directions in function
space: time--evolution (as a deformation in direction of the
extrinsic curvature of the space sections encoding the kinematical
variables), and scale--``evolution'' (as a deformation in
direction of the intrinsic 3--Ricci curvature). With regard to a
proper relation of those averages to observations, however, the
possibility of averaging on the lightcone has to be seriously
considered \cite{ellsto87}.

Employing such a logical split, it is reasonable to ask whether
the universe described by a kinematically averaged model
accelerates, independently of the question of whether {\it we
think} that the universe accelerates because we may be using the
wrong fitting model. With regard to the dark energy problem, the
question of whether a cosmological constant is needed in the
standard fitting model is then related to all of the effects
mentioned above, while the question addressed to the realistic
model only depends on the quantitative importance of the
kinematical backreaction compared with the other averaged sources
\cite{futamase2,bks,ras05,kolbetal05}. It should be emphasized that in Newtonian
cosmology global kinematical backreaction is absent, since (for
Euclidean space sections) the fluctuating source term is a total
divergence and thus vanishes for the periodic--boundary
architecture of Newtonian models \cite{buchertehlers}.

The key--issue is whether these backreaction effects are significant
in cosmology.\  On the one hand, they might play a significant role
in the inflationary era \cite{mukhanovetal,gesbra05}. On the other, it can
possibly help explain the apparent dynamical existence either of
dark energy and/or of dark matter as effective terms in the
macroscopic dynamics at recent times. Various papers suggest the
effect may indeed be significant, for example the observed
acceleration of the universe could possibly be the result of the
backreaction of cosmological perturbations rather than the effect
of a negative--pressure dark energy fluid \cite{wet03},
\cite {kolbetal05}. However, other studies obtain
different results \cite{russetal}, \cite{bks},  \cite{nam02},
\cite{not05}, \cite{ras05}. Gauge
effects are problematic \cite{bra02,gesbra02}, and many astrophysicists
doubt the effect is significant. The issue still has to be
resolved. In any case, a detailed investigation of backreaction effects helps to improve
fitting models on regional scales for a better interpretation of observational data.

\section{Entropy and coarse--graining}

Related to all this is the puzzling question of gravitational entropy. The
spontaneous structure growth in the expanding universe due to gravitational
attraction appears to be contrary to all the statements about entropy in
standard textbooks \cite{ell95}. This must somehow be related to the nature
of the entropy of the gravitational field itself, not just the entropy of
matter in a gravitational field.

Now the key--feature regarding the entropy of matter, as clearly
explained by Penrose \cite{pen89,pen04}, is that it is associated
with the loss of information that occurs with any coarse--grained
description of matter. The most likely macroscopic states will be
those that correspond to the largest numbers of microscopic
states; that is to the largest volumes of phase space. This is
made clear in Boltzmann's definition of entropy:\ $S\,=\,k\ln V_{\Gamma}$ where
$k$ is Boltzmann's constant and $V_{\Gamma}$ the volume of phase space with
points indistinguishable from each other by means of macroscopic
observations of some macro (coarse--grained) variable to some
accuracy $\varepsilon$. The dynamics of the system is accompanied by an increase of
this entropy as the representative point in phase space
moves from less probable to more probable states.

One might therefore expect that a proper definition of
gravitational entropy would similarly be related to some kind of
coarse--graining of the gravitational field. However, most
attempts at definitions of gravitational entropy in the
cosmological context (e.g. \cite{pelcol04}, \cite{am05}) build on
Penrose's proposal \cite{pen89,pen04} that it be related to the
magnitude of the Weyl tensor, with no introduction of coarse--graining.
This is quite puzzling, given the persuasiveness of
Penrose' arguments that in the case of matter descriptions,
entropy is always related to such coarse--graining. In our view
this is one of the most fundamental missing aspects of
gravitational theory: a satisfactory relation of gravitational
entropy for a general gravitational field in terms of a
coarse--grained description of that field, therefore relating to
all the issues mentioned in the preceding sections.

A promising start has been made by Hosoya et al. \cite{hosetal04}:
if we are only concerned with averaging the matter inhomogeneities on an
inhomogeneous geometry, one can {\it deduce} an entropy measure for the
distinguishability of the density distribution from its average value directly
from the non--commutativity rule:
\begin{equation}
\label{relativeentropy}
\partial_t \langle \varrho\rangle - \langle {\partial_t \varrho}\rangle \;=\;
 -\frac{1}{V}\;\partial_t {S}\lbrace\varrho || \langle\varrho\rangle\rbrace \quad;\quad
{S}\lbrace\varrho || \langle\varrho\rangle\rbrace:=
\int \varrho \ln \frac{\varrho}{\langle\varrho\rangle}\;\;,
\end{equation}
where the functional ${S}\lbrace\varrho ||
\langle\varrho\rangle\rbrace$ is known in information theory as
the {\it Kullback--Leibler relative entropy}, spatial averaging
and integration is performed over a domain with volume $V$.

 The conjecture has been made \cite{hosetal04} that this functional is, after
a sufficient period of time, always globally increasing. This (in
view of canonical considerations, e.g. in isolated Markovian
systems) counter--intuitive statement is justified in a
self--gravitating system because gravity is long--range, the
averaging domain is not isolated, and gravity invokes a negative
feedback: structural inhomogeneities are amplified due to
gravitational instability. We may expect that the information
content in the matter inhomogeneities is always increasing.

Given such a definition, the problem is to determine whether
increasing {\it total} entropy (in the gravitational field and in the matter distribution)
occurs always, or whether this is true only for special initial conditions.
As discussed by Penrose \cite{pen89,pen04}, it seems plausible that the latter is the
case, with the arrow of time in physics
arising from boundary conditions at the start and end of the
universe:\ specifically, the Weyl tensor taking a special form at
the start of the expansion of the universe but a generic form at
the end. The specific details of this proposal have never been clarified,
and it is possible that the relation is not due to the Weyl tensor {\it per se}
but rather due to a spatial integral of the divergence of the Electric part of the
Weyl tensor (Ellis and Tavakol, unpublished).
A further problem is then relating the arrow of time for
structure growth in the universe to that for electromagnetic and
gravitational radiation \cite{ellsci72}. Here again coarse--graining
is crucial, for this relates to the kind of multi--scale
description of the gravitational field envisaged by Isaacson, as
discussed above.

Entropy and the associated arrow of time are fundamental to
macroscopic physics. Their foundations in relation to microphysics
remain mysterious in the case of general gravitational fields. The
entropy of black holes is of course well understood, but this is
an extreme case that does not by itself help us understand the
relation of entropy to spontaneous structure formation in the expanding
universe. Until this is solved, we cannot claim to properly
understand the nature of entropy in the cosmological context \cite{ell02}.

\smallskip

{\it GE acknowledges support by the NRF, and TB by the
Sonderforschungsbereich SFB 375 `Astroparticle physics' of the
German Science Foundation DFG.}

\end{document}